\documentclass{revtex4}

\usepackage{graphicx}
\begin{document}

\thispagestyle{empty}
\setcounter{page}{0}
\def\thefootnote{\fnsymbol{footnote}}

\begin{flushright}
DESY 01-183 \\
Edinburgh 2001/18\\
KA-TP-31-2001 \\
UB-HET-01-08\\
hep-ph/0111060\\
\end{flushright}

\vspace{1cm}

\begin{center}

{\Large\sc {\bf Status and Prospects of Theoretical Predictions for Weak Gauge
}} 
\vspace*{0.4cm} 

{\Large\sc {\bf Boson Production Processes at Lepton and Hadron Colliders$^*$}}
 
\vspace{1cm}

{\sc U.~Baur} \\
\vspace*{0.2cm} 
State University of New York at Buffalo, Buffalo, NY 14260, USA \\

E-mail: baur@ubhex.physics.buffalo.edu \\

\vspace*{0.4cm} 

{\sc A.~Denner} \\
\vspace*{0.2cm} 
Paul Scherrer Institut, CH-5232 Villigen PSI, Switzerland \\

E-mail:Ansgar.Denner@psi.ch \\

\vspace*{0.4cm} 

{\sc S.~Dittmaier} \\
\vspace*{0.2cm} 
DESY, D-22603 Hamburg, Germany  \\

E-mail: Stefan.Dittmaier@desy.de \\

\vspace*{0.4cm} 

{\sc M.~Kr\"amer} \\
\vspace*{0.2cm} 
The University of Edinburgh, EH9 3JZ Edinburgh, Scotland \\

E-mail: Michael.Kraemer@ed.ac.uk

\vspace*{0.4cm} 

{\sc M.~Roth} \\
\vspace*{0.2cm} 
Universit\"at Karlsruhe, D-76128 Karlsruhe, Germany \\

E-mail: roth@particle.uni-karlsruhe.de \\

\vspace*{0.4cm} 

{\sc D.~Wackeroth} \\
\vspace*{0.2cm} 
University of Rochester, Rochester, NY 14627, USA \\

E-mail: dow@pas.rochester.edu 

\end{center}

\vspace*{1cm}

\begin{center}
{\large\bf Abstract}
\end{center}

For the envisioned precision measurement of the $W$-boson mass at
present and future lepton and hadron colliders it is crucial that the
theoretical predictions for the underlying production processes are
well under control.  We briefly describe the status of the predictions
for the $W$-pair-production processes at $e^+ e^-$ colliders, $e^+ e^-
\to W^+W^- \to 4f$, and for $W$- and $Z$-boson production at $pp$ and
$p\bar p$ colliders, $p\,p\hskip-7pt\hbox{$^{^{(\!-\!)}}$} \to W^{\pm}
\to \ell^{\pm} \nu_{\ell}$ and
$p\,p\hskip-7pt\hbox{$^{^{(\!-\!)}}$}\to Z,\gamma \to \ell^+ \ell^-$
($\ell=e,\,\mu$). We also discuss the theoretical improvements needed
to meet the experimental accuracies one hopes to achieve in future
experiments.
\vfill

$^*$ Contribution to the Proceedings of the ``Workshop on the Future
of Particle Physics'' (Snowmass 2001), Snowmass Village, Colorado,
USA, June 30 -- July 21, 2001.

\newpage

\bibliographystyle{revtex}

\title{Status and Prospects of Theoretical Predictions for Weak Gauge
Boson Production Processes at Lepton and Hadron Colliders}



\author{U.~Baur}
\email[]{baur@ubhex.physics.buffalo.edu}
\affiliation{State University of New York at Buffalo, Buffalo, NY 14260,
USA}

\author{A.~Denner}
\email[]{Ansgar.Denner@psi.ch}
\affiliation{Paul Scherrer Institut, CH-5232 Villigen PSI, Switzerland}

\author{S.~Dittmaier}
\email[]{Stefan.Dittmaier@desy.de}
\affiliation{DESY, D-22603 Hamburg, Germany}

\author{M.~Kr\"amer}
\email[]{Michael.Kraemer@ed.ac.uk}
\affiliation{The University of Edinburgh, EH9 3JZ Edinburgh, Scotland}

\author{M.~Roth}
\email[]{roth@particle.uni-karlsruhe.de}
\affiliation{Universit\"at Karlsruhe, D-76128 Karlsruhe, Germany}

\author{D.~Wackeroth}
\email[]{dow@pas.rochester.edu}
\affiliation{University of Rochester, Rochester, NY 14627, USA}


\date{\today}

\begin{abstract}
For the envisioned precision measurement of the $W$-boson mass at
present and future lepton and hadron colliders it is crucial that the
theoretical predictions for the underlying production processes are
well under control.  We briefly describe the status of the predictions
for the $W$-pair-production processes at $e^+ e^-$ colliders, $e^+ e^-
\to W^+W^- \to 4f$, and for $W$- and $Z$-boson production at $pp$ and
$p\bar p$ colliders, $p\,p\hskip-7pt\hbox{$^{^{(\!-\!)}}$} \to W^{\pm}
\to \ell^{\pm} \nu_{\ell}$ and
$p\,p\hskip-7pt\hbox{$^{^{(\!-\!)}}$}\to Z,\gamma \to \ell^+ \ell^-$
($\ell=e,\,\mu$). We also discuss the theoretical improvements needed
to meet the experimental accuracies one hopes to achieve in future
experiments.
\end{abstract}

\maketitle

\section{Introduction}
\label{sec:intro}

One of the main tasks at present and future lepton and hadron
colliders is a precise measurement of the $W$-boson mass, $M_W$.  The
present value of the $W$-boson mass, $M_W=80.451\pm 0.033$ GeV
\cite{lepewwg}, has been obtained from combining measurements carried
out at LEP2, the CERN $\bar pp$ collider, and in Run~I of the
Tevatron.  The $W$-boson mass plays a major role in the indirect
determination of the Standard Model Higgs-boson mass, $M_H$, from a
global fit to the electroweak precision observables.  A more precise
knowledge of $M_W$ will greatly improve the indirect bounds on $M_H$
as discussed in detail, for example, in Ref.~\cite{BlueBand}.

In future experiments at the Tevatron, one expects to measure $M_W$
with a precision of 20--40~MeV~\cite{run2prec}. At the LHC one hopes
to achieve an accuracy of 15~MeV~\cite{haywood}. At a future linear
$e^+e^-$ collider (LC) a precision of about 6
MeV~\cite{wilson,TeslaTDR} from a dedicated threshold scan of the
$W$-pair production cross section and of the order of
15~MeV~\cite{Accomando:1998wt} or even better~\cite{moenig} at
$\sqrt{s}=500$~GeV may be achievable.  In order to measure the
$W$-boson mass with such high precision it is necessary to fully
understand and control QCD and electroweak corrections to single-$W$
production at hadron colliders and to $W$-pair production at lepton
colliders.

At a linear collider, one expects that about $10^6$ $W$-boson pairs
are produced per year, compared to a yield of ${\cal O}(10^4)$
$W$-boson pairs at LEP2. This implies that the $e^+ e^- \to W^+W^- \to
4f$ cross section can be measured with a precision of the order of a
few per mille. The measurement of the $W$-pair differential cross
section provides a tool for a precise measurement of the (charged)
triple gauge-boson couplings (TGCs)~\cite{lepwwac}. Already at LEP2
the precision achieved for the TGCs requires the inclusion of
electroweak ${\cal O}(\alpha)$ corrections in the calculation of the
cross section. They affect the shape of the angular distribution from
which the TGCs are extracted and, in case of the $C$ and $P$
conserving TGCs, may introduce theoretical uncertainties as large as
the experimental uncertainties~\cite{aleph,markus}.  Thus, for the
precision envisioned at a LC~\cite{TeslaTDR,orangebook} the inclusion
of electroweak 1-loop corrections is indispensable.

Here we give a status report of radiative corrections to $e^+ e^- \to
W^+W^- \to 4f$ and to weak gauge-boson production at hadron
colliders. In Section~\ref{sec:ww} we briefly describe the presently
available calculations of electroweak corrections to $W$-pair
production at LEP2 and LC center-of-mass (CM) energies, and discuss
their uncertainties and the theoretical challenges for the envisioned
precision of $M_W$ at a future LC.  In Section~\ref{sec:had} we
describe the status of calculations of radiative corrections to $W$-
and $Z$-boson production at hadron colliders. QCD corrections only
indirectly influence the $W$-mass determination~\cite{run2prec} and
their effect on the $p_T$ distribution of the $W$~boson is well
described by calculations~\cite{resbos,resbos2,resbos3} which resum
soft-gluon emission terms. We therefore focus entirely on electroweak
radiative corrections in our discussion.  In Section~\ref{sec:theory}
we, finally, briefly discuss the field-theoretical definition of the
$W$-boson mass. The precise definition of $M_W$ is important at CM
energies much larger than the 
gauge-boson masses, and needs to be addressed for precision
measurements of $M_W$ at future $e^+ e^-$ and hadron colliders.

\section{Electroweak Radiative Corrections to \boldmath$e^+e^-\to WW\to 4f$}
\label{sec:ww}

A complete calculation of the ${\cal O}(\alpha)$ electroweak
corrections to $e^+e^-\to WW\to 4f$ is technically very difficult.
Moreover, the requirement to include finite $W$-width effects poses
severe problems with gauge invariance. Although there is ongoing work
in this direction~\cite{Vicini:1998iy,vicini}, a full calculation of
the ${\cal O}(\alpha)$ electroweak corrections to $e^+e^-\to WW\to 4f$
is currently not available. A suitable approach to include ${\cal
O}(\alpha)$ corrections to $W$-pair production is provided by the
double-pole approximation (DPA): electroweak ${\cal O}(\alpha)$
corrections are only considered for the terms that are enhanced by two
resonant $W$ bosons. The intrinsic DPA uncertainty is estimated to be
of the order of $\alpha \Gamma_W/(\pi M_W)$, i.e.\
${\mathrel{\raisebox{-.3em}{$\stackrel{\displaystyle <}{\sim}$}}} 0.5
\%$, whenever the cross section is dominated by doubly-resonant
contributions. This is the case at LEP2 energies sufficiently above
threshold and up to about 500 GeV.  The DPA is not a valid
approximation close to the $W$-pair production threshold.  At higher
energies contributions from single resonant (single $W$ production)
and non-resonant diagrams become sizeable, and appropriate cuts have
to be imposed to extract the $WW$ signal.  All present calculations of
${\cal O}(\alpha)$ corrections to $e^+e^-\to WW \to 4f$ rely on the
DPA~\cite{Beenakker:1999gr,Denner:2000bj,Jadach:1998hi,Jadach:2000tz,
Kurihara:2000ii}. However, the technical details of the DPA
implementation differ somewhat and lead to small differences in the
predicted cross sections within the expected uncertainties. These
differences can be studied in detail with the state-of-the-art Monte
Carlo (MC) generators {\tt
RacoonWW}~\cite{Denner:1999dt,Denner:1999kn,Denner:2000bj} and {\tt
YFSWW3}~\cite{Jadach:1998hi,Jadach:2000tz,Jadach:2001uu}, and have to
be treated as systematic theoretical uncertainties in the extraction
of $M_W$ and TGCs. In the following, we briefly discuss the
theoretical uncertainties associated with those MC generators, and the
theoretical precision required in order to meet the goal of $\delta
M_W=6$~MeV in a threshold scan at a LC.  A detailed comparison of all
presently available calculations for $e^+e^- \to 4f$ can be found in
Ref.~\cite{mclep2}.

\subsection{Theoretical Uncertainties of Current Calculations of
\boldmath $e^+e^-\to WW\to 4f$} 

A tuned numerical comparison between the state-of-the-art MC
generators {\tt RacoonWW} and {\tt YFSWW3}, supported by a comparison
with a semi-analytical calculation~\cite{Beenakker:1999gr} and a study
of the intrinsic DPA ambiguity with {\tt
RacoonWW}~\cite{mclep2,Denner:2000bj} and {\tt YFSWW3}~\cite{mclep2},
shows that the current theoretical uncertainty for the total $W$-pair
production cross section is about 0.5\% for CM energies
between 170~GeV and 500~GeV~\cite{mclep2}. This is in agreement with
the expected intrinsic DPA uncertainty of
${\mathrel{\raisebox{-.3em}{$\stackrel{\displaystyle <}{\sim}$}}} 0.5
\%$ for these energies.  A tuned comparison has also been performed
for {\tt RacoonWW} and {\tt YFSWW3} predictions for the $W$
invariant-mass distribution, the distribution of the $W$ production
angle, as well as several photon observables at
$\sqrt{s}=200$~GeV~\cite{mclep2,Denner:2001vr,Jadach:2000kw} and
$\sqrt{s}=500$~GeV~\cite{Denner:2000cg,Denner:2001vr,Jadach:2000kw}.
Taking the observed differences~\cite{mclep2} between the {\tt
RacoonWW} and {\tt YFSWW3} predictions as a guideline, a theoretical
uncertainty of about $1\%$ can be assigned to the distribution of the
$W$ production angle and the $W$ invariant-mass distribution in the
$W$ resonance region.

The theoretical uncertainties of the $e^+e^-\to WW\to 4f$ cross
section translate into uncertainties of the $W$ mass and the TGCs
extracted from data.  A recent study~\cite{Jadach:2001cz} based on the
MC generators {\tt KoralW} and {\tt YFSWW3} finds a theoretical
uncertainty of $\delta M_W=5$~MeV due to unknown electroweak
corrections at LEP2 energies. This is consistent with a qualitative
\lq naive\rq~estimate for the shift in $M_W$ derived from the observed
difference \cite{mclep2} between the {\tt RacoonWW} and {\tt YFSWW3}
predictions for the $W$ invariant-mass distribution.  Using the MC
generator {\tt YFSWW3}, the ALEPH collaboration~\cite{aleph} has
derived (preliminary) results for the shifts in the extracted values
for the TGCs due to the inclusion of electroweak corrections.  A study
of the theoretical uncertainty induced by
missing electroweak radiative corrections for the TGC parameter 
$\lambda=\lambda_{\gamma}=\lambda_Z$, extracted from the $W$ angular
distribution at LEP2, is in progress~\cite{YFSRAC}.

A major difference between {\tt RacoonWW} and {\tt YFSWW3} lies in
their treatment of visible photons. In {\tt RacoonWW} real photon
radiation is based on the full $4f+\gamma$ matrix
element and collinear leading higher-order initial-state radiation up
to ${\cal O}(\alpha^3)$. In
{\tt YFSWW3}, on the other hand, multi-photon radiation to $W$-pair
production is combined with ${\cal O}(\alpha^2)$ leading-logarithmic  
photon radiation
in $W$ decays by using {\tt PHOTOS}~\cite{Barberio:1991ms}.  A
comparison~\cite{Denner:2001vr} of {\tt RacoonWW} and {\tt YFSWW3}
predictions for photon observables finds relative differences of less
than $5\%$ at $\sqrt{s}=200$~GeV and about $10\%$ at
$\sqrt{s}=500$~GeV between the two generators.  Precise knowledge of
the photon observables will be needed for the measurement of photonic
quartic gauge-boson couplings in $4f+\gamma$ production at a future
LC. Even for the LEP2 measurement of these couplings~\cite{lepwwac}, an
improvement in the current theoretical prediction for the
photon-energy spectrum is desirable.

\subsection{\boldmath $M_W$ Measurement at a Future Linear Collider}
\label{subsec:mwlinac}

The $W$-boson mass can be measured in $W$-pair production at a LC
either in a dedicated threshold scan operating the machine at
$\sqrt{s}\approx 161$~GeV, or via direct reconstruction of the $W$
bosons at intermediate and high energies $\sqrt{s}=170$--1500~GeV.
Both strategies have been used with success at LEP2.

In the threshold region, the total $W$-pair production cross section,
$\sigma_{WW}$, is very sensitive to the $W$-boson mass. A rough
estimate of the statistical power of a $W$-mass measurement from
$\sigma_{WW}$ in this energy region at a LC can be obtained from the
corresponding LEP2 studies. As discussed in
Refs.~\cite{lep2rep,stirling_1} the sensitivity for $M_W$ is largest
in the region around $\sqrt{s}=161$~GeV at which point the statistical
uncertainty can be estimated to be
\begin{equation}
\delta M_W^{\rm stat}\approx 90~{\rm MeV}\left[{\varepsilon\int\!{\cal 
L}dt\over
100~{\rm pb}^{-1}}\right]^{-1/2}.
\label{eq:stat}
\end{equation}
Here, $\varepsilon$ is the efficiency for detecting $W$ bosons. For
$\varepsilon=0.67$ and an integrated luminosity of 100~fb$^{-1}$, one
finds from Eq.~(\ref{eq:stat})
\begin{equation}
\delta M_W^{\rm stat}\approx 3.5~{\rm MeV}.
\end{equation}
The systematic uncertainty due to an overall multiplicative factor $C$
to the cross section, such as efficiency and luminosity, can be
parametrized in the form~\cite{lep2rep,stirling_1},
\begin{equation}
\delta M_W^{\rm syst}=17~ {\rm MeV} \left[ \frac{\Delta C}{C}\times 100\right]  \, ,
\end{equation}
where $\Delta C/C$ is the relative error on the quantity
$C$.  Assuming that the efficiency and the integrated luminosity can
be determined with a precision of
$\Delta\varepsilon/\varepsilon=0.25\%$ and $\Delta{\cal L}/ {\cal
L}=0.1\%$, one finds that $M_W$ can be measured with an uncertainty of
\begin{equation}
\delta M_W\approx 6~{\rm MeV},
\end{equation}
provided that the theoretical uncertainty is under control in the
energy region of interest.  This estimate has been confirmed by a more
recent study~\cite{wilson} of the precision of a $M_W$ measurement
from a threshold scan at a LC using simulated data and taking into
account experimental systematic uncertainties. However, as emphasized
in Ref.~\cite{wilson}, for such a precise $M_W$ measurement not to be
jeopardized by the theoretical uncertainty (i.e.\ to keep the
theoretical uncertainty below 1--2 MeV) the calculations of the
relevant observables in the energy region of interest need to be of a
precision of ${\cal O}(0.1\%)$. 

Presently, the total $W$-pair production cross section in the
threshold region is known with an accuracy of about
1--2\%~\cite{Beenakker:1996kt,Denner:2001zp}, since predictions are
based on an improved-Born approximation which neglects non-universal
electroweak corrections.  As mentioned earlier, the DPA is not a valid
approximation in the threshold region, as the $e^+e^-\to 4f$ cross
section in this region is not dominated by $W$-pair production.  At
present there is no study available on how the requirement that the
theoretical uncertainty of $M_W$ extracted from a threshold scan has
to be less than 1--2 MeV translates into a constraint on the
theoretical precision for $\sigma_{WW}$ in the threshold region.  One
could argue that only the shape of the total $W$-pair production cross
section in the threshold region needs to be known with high precision
but not the overall normalization. However, it is expected that the
non-universal ${\cal O}(\alpha)$ corrections modify the shape of the
cross section curve in this energy region.  If one (pessimistically)
assumes that the theoretical uncertainty of the cross section will not
improve and that the shape of $\sigma_{WW}$ in the narrow region of
$\sqrt{s}\approx 161$ GeV considered in Ref.~\cite{wilson} is not
predicted with sufficient accuracy by the improved-Born approximation,
the uncertainty of the $W$ mass obtained from a threshold scan is
completely dominated by the theoretical error, and the precision of
the $W$ mass is limited to~\cite{stirling_1}
\begin{eqnarray}
\delta M_W \approx \delta M_W^{\rm theor} & \approx & 17~{\rm
MeV}\left[{\Delta\sigma\over\sigma}\times 100\right] 
                                    \approx 20\mbox{--}30~{\rm MeV}.
\end{eqnarray}
Thus, in order to understand and reduce the theoretical uncertainty of
the $M_W$ measurement from a threshold scan to the desired level, the
full ${\cal O}(\alpha)$ electroweak corrections to $e^+e^-\to 4f$ in
the threshold region are needed.

The full treatment of the processes $e^+e^- \to 4f$ at the one-loop
level is of enormous complexity.  While the real Bremsstrahlung
contribution, $e^+e^-\to 4f+\gamma$, is known exactly for all final
states~\cite{Caravaglios:1997nq,Denner:1999gp,Papadopoulos:2000tt,Jegerlehner:2001ki}
there are severe theoretical problems with the virtual ${\cal
O}(\alpha)$ corrections.  A full calculation of the virtual
contribution to $e^+e^-\to 4f$ can lead to ${\cal O}(10^4)$
complicated one-loop diagrams involving 5- and 6-point functions which
exhibit potential numerical instabilities.  Besides these technical
challenges, there are serious theoretical problems with gauge
invariance in connection with the instability of the $W$ and $Z$
bosons, as discussed in Section~\ref{sec:theory}.

Using direct reconstruction of $W$ bosons and assuming an integrated
luminosity of 500~fb$^{-1}$ at $\sqrt{s}=500$~GeV, one expects a
statistical error of $\delta M_W^{\rm stat}\approx 3.5~{\rm
MeV}$~\cite{moenig}. Systematic errors are dominated by jet resolution
effects. Using $Z\gamma$ events where the $Z$ decays into 2~jets and
the photon is lost in the beam pipe for calibration, a systematic
error $\delta M_W^{\rm syst}< 10$--15~{\rm MeV} is expected. The
resulting overall precision of the $W$-boson mass from direct $W$
reconstruction at a LC operating at an energy well above the $W$-pair
threshold is~\cite{TeslaTDR,moenig}
\begin{equation}
\delta M_W\approx 10\mbox{--}15~{\rm MeV}. 
\end{equation}
For theoretical predictions in the high-energy range, i.e.\ $\sqrt{s}>
500$ GeV, the same calculations as for the LEP2-energy range can be
used as a starting point. However, in view of a 10--15 MeV precision,
the study of the theoretical uncertainty of $M_W$ due to electroweak
corrections of Ref.~\cite{Jadach:2001cz} should be repeated at 500
GeV. Moreover, large electroweak logarithms of Sudakov-type become
increasingly important and the theoretical uncertainty is expected to
become worse due to missing higher-order corrections. At $\sqrt{s}=1$
TeV the typical size of the corrections from ${\cal O}(\alpha^2)$
Sudakov logarithms to cross sections for single-$W$ and $W$-pair
production processes can be estimated to be of the order of one and
several per cent, respectively. These double logarithmic contributions
are expected to exponentiate, which has been confirmed by explicit
two-loop calculations
\cite{Melles:2000ed,Hori:2000tm,Beenakker:2000kb}.  Subleading
contributions are also under study for specific processes, since they
are expected to have large coefficients at one loop.  For the relevant
literature and a review of the present understanding of electroweak
Sudakov logarithms we refer to
Refs.~\cite{Melles:2001ye,Denner:2001mn}.  The effect of these
corrections on the $W$ mass extracted from data has not been studied
yet.


\section{Electroweak Radiative Corrections to Weak Boson
Production in Hadronic Collisions}
\label{sec:had}

The determination of the $W$-boson mass in a hadron collider
environment requires a simultaneous precision measurement of the
$Z$-boson mass, $M_Z$, and width, $\Gamma_Z$. When compared to the
value measured at LEP1, the two quantities help to accurately
calibrate detector
components~\cite{cdfwmass,d0wmass,cdfwmass2,d0wmass2}. It is therefore
necessary to understand electroweak corrections for both $W$- and
$Z$-boson production.

QED corrections are known to produce a considerable shift in the
measured $W$- and $Z$-boson
masses~\cite{cdfwmass,d0wmass,cdfwmass2,d0wmass2}. The shift in $M_W$
is significantly larger than the uncertainty expected in future
experiments.  In the calculation~\cite{BK} which was used in the
analysis of the Tevatron Run~I data, only the final-state photonic
corrections were correctly included. The sum of the soft and virtual
parts was indirectly estimated from the inclusive ${\cal O}(\alpha^2)$
$Z\to\ell^+\ell^-(\gamma)$ and $W\to\ell\nu_\ell(\gamma)$ width and the
hard photon bremsstrahlung contribution. Initial-state, interference,
and weak contributions to the ${\cal O}(\alpha)$ corrections were
ignored altogether. The ignored parts of the ${\cal O}(\alpha)$
electroweak radiative corrections, combined with effects of multiple
photon emission (higher-order corrections), have been estimated to
contribute a systematic uncertainty of $\delta M_W= 15$--20~MeV to the
measurement of the $W$-boson
mass~\cite{cdfwmass,d0wmass,cdfwmass2,d0wmass2}. Given the expected
accuracy for $M_W$ in Run~II of the Tevatron and the LHC, improved
calculations of the electroweak corrections to weak-boson production
in hadronic collisions are needed.

The full electroweak ${\cal O}(\alpha)$ corrections to resonant
$W$-boson production in a general four-fermion process were computed
in Ref.~\cite{Wackeroth:1997hz} with special emphasis on obtaining a
gauge-invariant decomposition into a photonic and non-photonic
part. The results were used in Ref.~\cite{BKW} to calculate the ${\cal
O}(\alpha)$ electroweak corrections to
$p\,p\hskip-7pt\hbox{$^{^{(\!-\!)}}$} \to W\to\ell\nu_\ell$ in the pole
approximation. The cross section for $W$-boson production via the
Drell--Yan mechanism at parton level, $q_i \overline{q}_{i'}\rightarrow
f \bar{f}'(\gamma)$, in the pole approximation can be written in the
form
\begin{eqnarray}\label{eqwack:one}
d \hat{\sigma}^{(0+1)} & = &
d \hat \sigma^{(0)}\; \Bigl\{1+ 2 {\cal R}e  \Bigl[\tilde F_{{\rm weak}}^{{\rm initial}}(\hat s=M_W^2)+
\tilde F_{{\rm weak}}^{{\rm final}}(\hat s=M_W^2) \Bigr]\Bigr\} 
\nonumber \\
&&+ \sum_{{\rm a}={\rm initial},{\rm final},\atop {\rm interf.}} 
 \Bigl[d\hat\sigma^{(0)}\; 
F_{{\rm QED}}^{{\rm a}}(\hat s,\hat t)+
d \hat \sigma_{2\rightarrow 3}^{{\rm a}} \Bigr] \; ,
\end{eqnarray}
where the Born-cross section, $d \hat \sigma^{(0)}$, is of
Breit--Wigner form, and $\hat s$ and $\hat t$ are the usual Mandelstam
variables in the parton CM frame.  The (modified) weak
corrections and the virtual and soft-photon emission from the initial
and final-state fermions (as well as their interference) are described
by the form factors $\tilde F_{{\rm weak}}^{{\rm a}}$ and $F_{{\rm
    QED}}^{{\rm a}}$, respectively.  The IR-finite contribution
$d\hat\sigma_{2\rightarrow 3}^{{\rm a}}$ describes real photon
radiation away from soft singularities.  Mass singularities of the
form $\ln(\hat s/m_f^2)$ arise when the photon is emitted collinear to
a charged fermion and the resulting singularity is regularized by
retaining a finite fermion mass ($m_f$).  $F_{{\rm QED}}^{{\rm
    initial}}$ and $d\hat\sigma_{2\rightarrow 3}^{{\rm initial}}$
still include quark-mass singularities which need to be extracted and
absorbed into the parton distribution functions (PDFs).  The
absorption of the quark-mass singularities into the PDFs can be done
in complete analogy to gluon emission in QCD, thereby introducing a
QED factorization scheme dependence.  The latter results in a modified
scale dependence of the PDFs, which is expected to have a negligible
effect on the observable cross
sections~\cite{Kripfganz:1988bd,Spiesberger:1995dm,haywood}.
Comparing the $W$-mass shifts obtained using the calculations of
Refs.~\cite{BK} and~\cite{BKW}, one finds that the proper treatment of
virtual and soft corrections and the inclusion of weak corrections
induces an additional shift of ${\cal O}$(10~MeV) in the extracted
$W$-boson mass.

The calculation of the electroweak ${\cal O}(\alpha)$ corrections to
$W$ production described in Ref.~\cite{BKW} was carried out in the
pole approximation, i.e.\ corrections which are very small at the $W$
pole, such as the $WZ$ box diagrams, were ignored and the form factors
$\tilde F_{{\rm weak}}^{{\rm a}}$ were evaluated at $\hat s=M_W^2$.  A
calculation of the full ${\cal O}(\alpha^3)$ matrix elements for
$p\,p\hskip-7pt\hbox{$^{^{(\!-\!)}}$} \rightarrow W\to\ell\nu_\ell$ 
has recently appeared in Refs.~\cite{dittm,Zykunov:2001mn}. An independent
calculation which will also discuss the impact of the electroweak
radiative corrections on the $M_W$ mass extracted from various
observables~\cite{run2prec} is in progress~\cite{BW}.
While the corrections ignored in Ref.~\cite{BKW} change the
differential cross section in the $W$-pole region by less than
1\%~\cite{dittm}, they become large at high $\ell\nu_\ell$ invariant
masses $m(\ell\nu_\ell)$ due to
Sudakov-like logarithms of the form $\ln^2[m(\ell\nu_\ell)/M_W]$.
They significantly affect the transverse mass distribution above the
$W$ peak, which serves as tool for a direct measurement of the $W$
width, $\Gamma_W$. Taking these corrections into account in future
measurements of $\Gamma_W$ will be important.

A calculation of the ${\cal O}(\alpha)$ QED corrections to
$p\,p\hskip-7pt\hbox{$^{^{(\!-\!)}}$} \to \gamma,\,Z\to\ell^+\ell^-$
based on the full set of contributing one-loop Feynman diagrams was
carried out in Ref.~\cite{BKS}. Purely weak corrections were ignored
in this first step towards a complete calculation of the ${\cal
  O}(\alpha)$ electroweak corrections to
$p\,p\hskip-7pt\hbox{$^{^{(\!-\!)}}$} \to \gamma,\,Z\to\ell^+\ell^-$.
The difference in the extracted $Z$-boson mass when comparing the
approximate calculation of Ref.~\cite{BK} with the full calculation of
the ${\cal O}(\alpha)$ QED corrections was found to be of ${\cal
  O}$(10~MeV). However, in order to properly calibrate the $Z$-boson
mass and width using the available LEP data, it is desirable to use
exactly the same theoretical input that has been used to extract
$M_Z$ and $\Gamma_Z$ at LEP, i.e.\ to include the purely weak
corrections to $p\,p\hskip-7pt\hbox{$^{^{(\!-\!)}}$} \to
\gamma,\,Z\to\ell^+\ell^-$ and the ${\cal O}(g^4m^2_t/M_W^2)$
corrections to the effective leptonic weak mixing parameter,
$\sin^2\theta_{\rm eff}^{\ell}$, and the $W$-boson mass~\cite{dg}, in the
calculation. A calculation of the complete ${\cal O}(\alpha)$
corrections to $p\,p\hskip-7pt\hbox{$^{^{(\!-\!)  }}$} \rightarrow
\gamma,\, Z \rightarrow \ell^+ \ell^-$ which also takes into account
the ${\cal O}(g^4m^2_t/M_W^2)$ corrections has recently been
completed~\cite{BBHSW}. The additional corrections taken into account
in Ref.~\cite{BBHSW} enhance the differential cross section in the
$Z$-peak region by up to 1.2\%. Since they are not uniform, they are
expected to shift the $Z$-boson mass extracted from data upward by
several MeV. Detailed simulations of this effect, however, have not
been carried out yet.

For the analysis of Run~IIA data~\cite{run2prec} an ${\cal
O}(\alpha^3)$ calculation of $W$ and $Z$ production will likely be
sufficient.  This is not the case for the precision of $M_W$ expected
with Run~IIB data ($\delta M_W\leq 20$~MeV). The main contribution to
the shift in $M_W$ originates from final-state photon radiation. An
explicit calculation of real two-photon radiation in $W$- and $Z$-boson
production~\cite{BS} indicates that, in order to measure the $W$ mass
with a precision of less than 20~MeV in a hadron collider environment
as foreseen in Run~IIB and at the LHC, it will be necessary to take
into account multi-photon radiation effects.

\section{Theoretical issues at high luminosity and energy}
\label{sec:theory}

For precision measurements of $M_W$ with values of $\delta M_W$
smaller than about 40~MeV, the precise definition of the $W$ mass and
width become important when these quantities are extracted.  For
stable particles the location of the pole in the particle's propagator
provides a proper definition of the mass parameter, which is known as the
pole mass. For unstable particles, such as the $W$ and $Z$ bosons, the
pole position becomes a complex quantity, so that the actual
definition of an associated real mass parameter involves some
convention. The imaginary part of the pole position is related to the
decay width of the unstable particle.  In LEP1 precision physics,
usually the convention, called the on-shell mass scheme, was adopted of
identifying the $Z$-boson mass with the zero in the real part of the
inverse propagator, which coincides with the real part of the complex pole
position in one-loop order, but differs at two loops and beyond.
Field-theoretical studies~\cite{Gambino:1999ai,Grassi:2001bz} showed
that the complex pole is related to the bare mass of the
unrenormalized field theory in a gauge-invariant way, while the
on-shell definition involves gauge dependences starting at two-loop
accuracy.

In practice the issue of the mass definition enters when the
Breit--Wigner resonance shape is parametrized by the mass and the
width. In the following we consider a single $W$-boson resonance, as
it appears, for instance, in $q\bar q'\to W\to\ell\nu_\ell$.  In a
field-theoretical description, the resonance propagator results from a
Dyson summation of the imaginary part of the $W$ vacuum polarization,
and the matrix element ${\cal M}$ behaves as
\begin{equation}
{\cal M} \sim \frac{R}{k^2-M_W^2+i M_W\Gamma_W(k^2)}
\label{eq:resonance}
\end{equation}
in the vicinity of the resonance ($k^2\sim M_W^2$), where $k$ is the
momentum transfer in the propagator.  This leads to an
energy-dependent width $\Gamma_W(k^2)$, which is given by
$\Gamma_W(k^2)=\Gamma_W\times (k^2/M_W^2)$ if the decay fermions are
taken to be massless. In Eq.~(\ref{eq:resonance}) we did not yet
specify the precise definition of the mass and width parameters.  In
the on-shell mass definition the form (\ref{eq:resonance}) naturally
arises in any loop order, thus we identify $M_W$ and $\Gamma_W$ with
the on-shell quantities in the following. The relation to the complex
pole position, denoted by
$\overline{M}_W^2-i\overline{M}_W\overline{\Gamma}_W$, is easily
obtained by inserting the explicit form of $\Gamma_W(k^2)$ into
Eq.~(\ref{eq:resonance}), and identifying terms. This leads to
\begin{equation}
{\cal M} \sim \frac{\overline{R}}
{k^2-\overline{M}_W^2+i \overline{M}_W\overline{\Gamma}_W}
\end{equation}
with \cite{bardin,Wackeroth:1997hz,Beenakker:1997kn} 
\begin{eqnarray}
\overline{M}_W &=& M_W/\sqrt{1+\gamma^2} 
= M_W-\frac{\Gamma_W^2}{2M_W} + \dots, 
\nonumber\\
\overline{\Gamma}_W &=& \Gamma_W/\sqrt{1+\gamma^2}
= \Gamma_W-\frac{\Gamma_W^3}{2M_W^2} + \dots, 
\nonumber\\
\overline{R} &=& R/(1+i\gamma)
= R\left(1-i\frac{\Gamma_W}{M_W} + \dots \right),
\label{eq:wparam}
\end{eqnarray}
where $\gamma=\Gamma_W/M_W$.  Thus, if the measured $W$~resonance is
fitted to a propagator that is parametrized with a constant width, the
measured mass parameter will be $\overline{M}_W$, which is about 27~MeV 
smaller than the on-shell value $M_W$.  This procedure ignores
distortion effects of the resonance shape, in particular those induced
by radiation.  In Ref.~\cite{dittm} it has been verified numerically
that the reparametrization (\ref{eq:wparam}) works extremely well
if electroweak ${\cal O}(\alpha)$ corrections are taken into account.

In the past, an energy-dependent $W$ width has been used in
measurements of the $W$ mass at the
Tevatron~\cite{cdfwmass,d0wmass,cdfwmass2,d0wmass2}. The Monte Carlo
programs available for the $W$-mass analysis at LEP2 (see
Ref.~\cite{mclep2} for an overview) in contrast use a constant $W$
width. Since the difference between the $W$ mass obtained using a
constant and an energy-dependent width is of the same size as the
current uncertainties from LEP2 and Tevatron data and significantly
larger than those expected from future collider experiments, it is
important to correct for this difference.

Unfortunately, the Dyson summation of the resonance propagators, as
described above, carries the risk of breaking gauge invariance. Gauge
invariance works order by order in perturbation theory. By resumming
the self-energy corrections, one only takes into account part of the
higher-order corrections. Apart from being theoretically questionable,
breaking gauge invariance may result in large numerical errors in
cross section calculations.  This has been illustrated by specific
examples in
Refs.~\cite{Argyres:1995ym,Beenakker:1997kn,Denner:1999gp}.

In order to restore gauge invariance, one can adopt the strategy of
finding the minimal set of Feynman diagrams that is necessary for
compensating the terms caused by an energy-dependent width that
violate gauge invariance.  In
Refs.~\cite{Baur:1995aa,Argyres:1995ym,Beenakker:1997kn,Passarino:1999zh,
Ballestrero:2000kk} this approach was systematically carried out by a
gauge-invariant Dyson summation of the fermion-loop corrections which
lead to the gauge-boson decay widths in the propagators.  This
procedure provides a consistent treatment of lowest-order matrix
elements, improved by running couplings.  For lowest-order predictions
there are also other methods leading to gauge-invariant results, such
as the complex-mass scheme \cite{Denner:1999gp} or the use of
effective Lagrangians \cite{Beenakker:1999hi}.  However, a practical
general way of combining Dyson-summed propagators with full ${\cal
O}(\alpha)$ corrections (including bosonic loops) is still not
available.  To this end, the background-field quantization seems to be
a promising framework, since Dyson summation preserves all Ward
identities induced by gauge invariance \cite{Denner:1996gb} in
contrast to conventional quantization.
  
For the resonance process $q\bar q'\to W\to \ell\nu_\ell$ the
gauge-invariant inclusion~\cite{Wackeroth:1997hz,dittm} of the ${\cal
O}(\alpha)$ correction was possible in a simple way, but this is due
to the occurrence of a single $s$-channel resonance.  While the naive
introduction of fixed gauge-boson widths might be sufficient for the
full calculation of the ${\cal O}(\alpha)$ corrections to the
four-fermion processes in the LEP2-energy range, the calculation of
these corrections at high energies certainly requires some progress
towards a solution of the gauge-invariance problem.

\section{Conclusions}
\label{sec:concl}

While the full electroweak ${\cal O}(\alpha)$ corrections to the
Drell--Yan-like production of $W$ and $Z$ bosons are known, all
calculations of ${\cal O}(\alpha)$ corrections to $e^+e^-\to WW \to
4f$ are based on the leading term of an expansion about the double
resonance pole, an approach that is known as double-pole
approximation.  For the $W$-mass measurements at LEP2 and in Run~IIA
($2~{\rm fb}^{-1}$) of the Tevatron, the accuracy of the corresponding
predictions is sufficient.

In order to measure the $W$-boson mass with a precision of less than
20~MeV in Run~IIB ($15~{\rm fb}^{-1}$) of the Tevatron and at the LHC,
radiative corrections beyond ${\cal O}(\alpha)$ have to be controlled.
In particular, multi-photon radiation effects~\cite{run2prec,BS} have
to be taken into account.

To fully exploit the potential of a future LC for a precision $M_W$
measurement in a threshold scan, the full ${\cal O}(\alpha)$
corrections to $e^+e^-\to 4f$ are needed.  On the other hand, for a
determination of $M_W$ by reconstruction at intermediate and high
energies ($\sqrt{s}=170 - 1500$~GeV), which is comparable in precision
with the measurements at Run~IIB and the LHC, the existing
calculations for radiative corrections to $e^+e^-\to WW\to 4f$ could
be sufficient, but this expectation needs to be confirmed by explicit
calculations, or proper estimates, of the missing corrections.


\begin{acknowledgments}
  One of us (U.B.) would like to thank the Phenomenology Institute of
  the University of Wisconsin, Madison, and the Fermilab Theory Group,
  where part of this work was carried out, for their splendid
  hospitality.  The work of U.B.  is supported by NSF grant
  PHY-9970703. The work of A.D. is supported in part by the Swiss
  Bundesamt f\"ur Bildung und Wissenschaft contract 99.0043 and the
  work of A.D., S.D., M.K., and M.R. by the European Union under
  contract HPRN-CT-2000-00149. S.D. is supported by a Heisenberg
  fellowship.  The work of D.W. is supported by the U.S.  Department
  of Energy under grant DE-FG02-91ER40685.
\end{acknowledgments}


\bibliography{snow_mw}

\begin{thebibliography}{10}
\providecommand*{\bibinfo}[2]{#2}
\providecommand*{\eprint}[1]{#1}
\providecommand*{\url}[1]{#1}
\bibitem{lepewwg}
The LEP Electroweak Working group, {\tt
  http://lepewwg.web.cern.ch/LEPEWWG/Welcome.html}.
\bibitem{BlueBand}
\bibinfo{author}{U.~Baur} \emph{et~al.}, contribution of the Working Group
  P1-WG1 to the proceedings of the {\it Workshop on the Future of Particle
  Physics} (Snowmass 2001), Snowmass Village, Colorado, USA, June 30 -- July
  21, 2001.
\bibitem{run2prec}
\bibinfo{author}{R.~Brock} \emph{et~al.}, \eprint{arXiv:hep-ex/0011009}.
\bibitem{haywood}
\bibinfo{author}{S.~Haywood} \emph{et~al.}, \eprint{arXiv:hep-ph/0003275}.
\bibitem{wilson}
\bibinfo{author}{G.~Wilson}, ~$\!\!\!$LC-PHSM-2001-009.
\bibitem{TeslaTDR}
\bibinfo{author}{J.~A. Aguilar-Saavedra} \emph{et~al.}
  (\bibinfo{collaboration}{ECFA/DESY LC Physics Working Group}), {TESLA}
  technical design report, \eprint{arXiv:hep-ph/0106315}.
\bibitem{Accomando:1998wt}
\bibinfo{author}{E.~Accomando} \emph{et~al.} (\bibinfo{collaboration}{ECFA/DESY
  LC Physics Working Group}), \bibinfo{journal}{Phys. Rept.}
  \bibinfo{volume}{\textbf{299}}, \bibinfo{pages}{1} (\bibinfo{date}{1998}),
  \eprint{arXiv:hep-ph/9705442}.
\bibitem{moenig}
\bibinfo{author}{K.~M{\"o}nig} and \bibinfo{author}{A.~Tonazzo}, talk at the
  Linear Collider Workshop, Padova, Italy, May 2000.
\bibitem{lepwwac}
The LEP Electroweak Gauge-Couplings Group, {\tt
  http://lepewwg.web.cern.ch/LEPEWWG/lepww/tgc}.
\bibitem{aleph}
\bibinfo{author}{{The ALEPH Collaboration}}, {ALEPH 2001-060, CONF 2001-040}.
\bibitem{markus}
\bibinfo{author}{A.~Denner}, \bibinfo{author}{S.~Dittmaier},
  \bibinfo{author}{M.~Roth}, and \bibinfo{author}{D.~Wackeroth},
  \eprint{arXiv:hep-ph/0110402}.
\bibitem{orangebook}
\bibinfo{author}{T.~Abe} \emph{et~al.} (\bibinfo{collaboration}{American Linear
  Collider Working Group}), {Linear} collider physics resource book for
  Snowmass 2001, \eprint{arXiv:hep-ex/0106057}.
\bibitem{resbos}
\bibinfo{author}{C.~Balazs}, \bibinfo{author}{J.~W. Qiu}, and
  \bibinfo{author}{C.~P. Yuan}, \bibinfo{journal}{Phys. Lett.}
  \bibinfo{volume}{\textbf{B355}}, \bibinfo{pages}{548} (\bibinfo{date}{1995}),
  \eprint{arXiv:hep-ph/9505203}.
\bibitem{resbos2}
\bibinfo{author}{R.~K. Ellis} and \bibinfo{author}{S.~Veseli},
  \bibinfo{journal}{Nucl. Phys.} \bibinfo{volume}{\textbf{B511}},
  \bibinfo{pages}{649} (\bibinfo{date}{1998}), \eprint{arXiv:hep-ph/9706526}.
\bibitem{resbos3}
\bibinfo{author}{A.~Kulesza} and \bibinfo{author}{W.~J. Stirling},
  \bibinfo{journal}{Eur. Phys. J.} \bibinfo{volume}{\textbf{C20}},
  \bibinfo{pages}{349} (\bibinfo{date}{2001}), \eprint{arXiv:hep-ph/0103089}.
\bibitem{Vicini:1998iy}
\bibinfo{author}{A.~Vicini}, \bibinfo{journal}{Acta Phys. Polon.}
  \bibinfo{volume}{\textbf{B29}}, \bibinfo{pages}{2847} (\bibinfo{date}{1998}).
\bibitem{vicini}
\bibinfo{author}{A.~Vicini}, \eprint{arXiv:hep-ph/0104164}.
\bibitem{Beenakker:1999gr}
\bibinfo{author}{W.~Beenakker}, \bibinfo{author}{F.~A. Berends}, and
  \bibinfo{author}{A.~P. Chapovsky}, \bibinfo{journal}{Nucl. Phys.}
  \bibinfo{volume}{\textbf{B548}}, \bibinfo{pages}{3} (\bibinfo{date}{1999}),
  \eprint{arXiv:hep-ph/9811481}.
\bibitem{Denner:2000bj}
\bibinfo{author}{A.~Denner}, \bibinfo{author}{S.~Dittmaier},
  \bibinfo{author}{M.~Roth}, and \bibinfo{author}{D.~Wackeroth},
  \bibinfo{journal}{Nucl. Phys.} \bibinfo{volume}{\textbf{B587}},
  \bibinfo{pages}{67} (\bibinfo{date}{2000}), \eprint{arXiv:hep-ph/0006307}.
\bibitem{Jadach:1998hi}
\bibinfo{author}{S.~Jadach}, \bibinfo{author}{W.~Placzek},
  \bibinfo{author}{M.~Skrzypek}, \bibinfo{author}{B.~F.~L. Ward}, and
  \bibinfo{author}{Z.~Was}, \bibinfo{journal}{Phys. Lett.}
  \bibinfo{volume}{\textbf{B417}}, \bibinfo{pages}{326} (\bibinfo{date}{1998}),
  \eprint{arXiv:hep-ph/9705429}.
\bibitem{Jadach:2000tz}
\bibinfo{author}{S.~Jadach}, \bibinfo{author}{W.~Placzek},
  \bibinfo{author}{M.~Skrzypek}, \bibinfo{author}{B.~F.~L. Ward}, and
  \bibinfo{author}{Z.~Was}, \bibinfo{journal}{Phys. Rev.}
  \bibinfo{volume}{\textbf{D61}}, \bibinfo{pages}{113010}
  (\bibinfo{date}{2000}), \eprint{arXiv:hep-ph/9907436}.
\bibitem{Kurihara:2000ii}
\bibinfo{author}{Y.~Kurihara}, \bibinfo{author}{M.~Kuroda}, and
  \bibinfo{author}{D.~Schildknecht}, \bibinfo{journal}{Nucl. Phys.}
  \bibinfo{volume}{\textbf{B565}}, \bibinfo{pages}{49} (\bibinfo{date}{2000}),
  \eprint{arXiv:hep-ph/9908486}.
\bibitem{Denner:1999dt}
\bibinfo{author}{A.~Denner}, \bibinfo{author}{S.~Dittmaier},
  \bibinfo{author}{M.~Roth}, and \bibinfo{author}{D.~Wackeroth},
  \bibinfo{journal}{Eur. Phys. J. direct} \bibinfo{volume}{\textbf{C4}},
  \bibinfo{pages}{1} (\bibinfo{date}{2000}), \eprint{arXiv:hep-ph/9912447}.
\bibitem{Denner:1999kn}
\bibinfo{author}{A.~Denner}, \bibinfo{author}{S.~Dittmaier},
  \bibinfo{author}{M.~Roth}, and \bibinfo{author}{D.~Wackeroth},
  \bibinfo{journal}{Phys. Lett.} \bibinfo{volume}{\textbf{B475}},
  \bibinfo{pages}{127} (\bibinfo{date}{2000}), \eprint{arXiv:hep-ph/9912261}.
\bibitem{Jadach:2001uu}
\bibinfo{author}{S.~Jadach}, \bibinfo{author}{W.~Placzek},
  \bibinfo{author}{M.~Skrzypek}, \bibinfo{author}{B.~F.~L. Ward}, and
  \bibinfo{author}{Z.~Was}, \bibinfo{journal}{Comput. Phys. Commun.}
  \bibinfo{volume}{\textbf{140}}, \bibinfo{pages}{432} (\bibinfo{date}{2001}),
  \eprint{arXiv:hep-ph/0103163}.
\bibitem{mclep2}
\bibinfo{author}{M.~W. Gr{\"u}newald} \emph{et~al.},
  \eprint{arXiv:hep-ph/0005309}.
\bibitem{Denner:2001vr}
\bibinfo{author}{A.~Denner}, \bibinfo{author}{S.~Dittmaier},
  \bibinfo{author}{M.~Roth}, and \bibinfo{author}{D.~Wackeroth},
  \bibinfo{journal}{Eur. Phys. J.} \bibinfo{volume}{\textbf{C20}},
  \bibinfo{pages}{201} (\bibinfo{date}{2001}), \eprint{arXiv:hep-ph/0104057}.
\bibitem{Jadach:2000kw}
\bibinfo{author}{S.~Jadach}, \bibinfo{author}{W.~Placzek},
  \bibinfo{author}{M.~Skrzypek}, \bibinfo{author}{B.~F.~L. Ward}, and
  \bibinfo{author}{Z.~Was}, \eprint{arXiv:hep-ph/0007012}.
\bibitem{Denner:2000cg}
\bibinfo{author}{A.~Denner}, \bibinfo{author}{S.~Dittmaier},
  \bibinfo{author}{M.~Roth}, and \bibinfo{author}{D.~Wackeroth},
  \eprint{arXiv:hep-ph/0007245}.
\bibitem{Jadach:2001cz}
\bibinfo{author}{S.~Jadach}, \bibinfo{author}{W.~Placzek},
  \bibinfo{author}{M.~Skrzypek}, \bibinfo{author}{B.~F.~L. Ward}, and
  \bibinfo{author}{Z.~Was}, \eprint{arXiv:hep-ph/0109072}.
\bibitem{YFSRAC}
\bibinfo{author}{R.~Bruneli\`ere} \emph{et~al.}, {CERN-TH/2001-274}, in
  preparation.
\bibitem{Barberio:1991ms}
\bibinfo{author}{E.~Barberio}, \bibinfo{author}{B.~van Eijk}, and
  \bibinfo{author}{Z.~Was}, \bibinfo{journal}{Comput. Phys. Commun.}
  \bibinfo{volume}{\textbf{66}}, \bibinfo{pages}{115} (\bibinfo{date}{1991}).
\bibitem{lep2rep}
\bibinfo{author}{Z.~Kunszt} \emph{et~al.}, in CERN 96-01, p.~141,
  \eprint{arXiv:hep-ph/9602352}.
\bibitem{stirling_1}
\bibinfo{author}{W.~J. Stirling}, \bibinfo{journal}{Nucl. Phys.}
  \bibinfo{volume}{\textbf{B456}}, \bibinfo{pages}{3} (\bibinfo{date}{1995}),
  \eprint{arXiv:hep-ph/9503320}.
\bibitem{Beenakker:1996kt}
\bibinfo{author}{W.~Beenakker} \emph{et~al.}, in CERN 96-01, p.~79,
  \eprint{arXiv:hep-ph/9602351}.
\bibitem{Denner:2001zp}
\bibinfo{author}{A.~Denner}, \bibinfo{author}{S.~Dittmaier},
  \bibinfo{author}{M.~Roth}, and \bibinfo{author}{D.~Wackeroth},
  \eprint{arXiv:hep-ph/0101257}.
\bibitem{Caravaglios:1997nq}
\bibinfo{author}{F.~Caravaglios} and \bibinfo{author}{M.~Moretti},
  \bibinfo{journal}{Z. Phys.} \bibinfo{volume}{\textbf{C74}},
  \bibinfo{pages}{291} (\bibinfo{date}{1997}), \eprint{arXiv:hep-ph/9604316}.
\bibitem{Denner:1999gp}
\bibinfo{author}{A.~Denner}, \bibinfo{author}{S.~Dittmaier},
  \bibinfo{author}{M.~Roth}, and \bibinfo{author}{D.~Wackeroth},
  \bibinfo{journal}{Nucl. Phys.} \bibinfo{volume}{\textbf{B560}},
  \bibinfo{pages}{33} (\bibinfo{date}{1999}), \eprint{arXiv:hep-ph/9904472}.
\bibitem{Papadopoulos:2000tt}
\bibinfo{author}{C.~G. Papadopoulos}, \bibinfo{journal}{Comput. Phys. Commun.}
  \bibinfo{volume}{\textbf{137}}, \bibinfo{pages}{247} (\bibinfo{date}{2001}),
  \eprint{arXiv:hep-ph/0007335}.
\bibitem{Jegerlehner:2001ki}
\bibinfo{author}{F.~Jegerlehner} and \bibinfo{author}{K.~Kolodziej}
  (\bibinfo{date}{2001}), \eprint{arXiv:hep-ph/0109290}.
\bibitem{Melles:2000ed}
\bibinfo{author}{M.~Melles}, \bibinfo{journal}{Phys. Lett.}
  \bibinfo{volume}{\textbf{B495}}, \bibinfo{pages}{81} (\bibinfo{date}{2000}),
  \eprint{arXiv:hep-ph/0006077}.
\bibitem{Hori:2000tm}
\bibinfo{author}{M.~Hori}, \bibinfo{author}{H.~Kawamura}, and
  \bibinfo{author}{J.~Kodaira}, \bibinfo{journal}{Phys. Lett.}
  \bibinfo{volume}{\textbf{B491}}, \bibinfo{pages}{275} (\bibinfo{date}{2000}),
  \eprint{arXiv:hep-ph/0007329}.
\bibitem{Beenakker:2000kb}
\bibinfo{author}{W.~Beenakker} and \bibinfo{author}{A.~Werthenbach},
  \bibinfo{journal}{Phys. Lett.} \bibinfo{volume}{\textbf{B489}},
  \bibinfo{pages}{148} (\bibinfo{date}{2000}), \eprint{arXiv:hep-ph/0005316}.
\bibitem{Melles:2001ye}
\bibinfo{author}{M.~Melles}, \eprint{arXiv:hep-ph/0104232}.
\bibitem{Denner:2001mn}
\bibinfo{author}{A.~Denner}, \eprint{arXiv:hep-ph/0110155}.
\bibitem{cdfwmass}
\bibinfo{author}{F.~Abe} \emph{et~al.} (\bibinfo{collaboration}{CDF
  Collaboration}), \bibinfo{journal}{Phys. Rev. Lett.}
  \bibinfo{volume}{\textbf{75}}, \bibinfo{pages}{11} (\bibinfo{date}{1995}),
  \eprint{arXiv:hep-ex/9503007}.
\bibitem{d0wmass}
\bibinfo{author}{S.~Abachi} \emph{et~al.} (\bibinfo{collaboration}{D{\O\ }
  Collaboration}), \bibinfo{journal}{Phys. Rev. Lett.}
  \bibinfo{volume}{\textbf{77}}, \bibinfo{pages}{3309} (\bibinfo{date}{1996}),
  \eprint{arXiv:hep-ex/9607011}.
\bibitem{cdfwmass2}
\bibinfo{author}{T.~Affolder} \emph{et~al.} (\bibinfo{collaboration}{CDF
  Collaboration}), \bibinfo{journal}{Phys. Rev.}
  \bibinfo{volume}{\textbf{D64}}, \bibinfo{pages}{052001}
  (\bibinfo{date}{2001}), \eprint{arXiv:hep-ex/0007044}.
\bibitem{d0wmass2}
\bibinfo{author}{B.~Abbott} \emph{et~al.} (\bibinfo{collaboration}{D{\O\ }
  Collaboration}), \bibinfo{journal}{Phys. Rev.}
  \bibinfo{volume}{\textbf{D62}}, \bibinfo{pages}{092006}
  (\bibinfo{date}{2000}), \eprint{arXiv:hep-ex/9908057}.
\bibitem{BK}
\bibinfo{author}{F.~A. Berends} and \bibinfo{author}{R.~Kleiss},
  \bibinfo{journal}{Z. Phys.} \bibinfo{volume}{\textbf{C27}},
  \bibinfo{pages}{365} (\bibinfo{date}{1985}).
\bibitem{Wackeroth:1997hz}
\bibinfo{author}{D.~Wackeroth} and \bibinfo{author}{W.~Hollik},
  \bibinfo{journal}{Phys. Rev.} \bibinfo{volume}{\textbf{D55}},
  \bibinfo{pages}{6788} (\bibinfo{date}{1997}), \eprint{arXiv:hep-ph/9606398}.
\bibitem{BKW}
\bibinfo{author}{U.~Baur}, \bibinfo{author}{S.~Keller}, and
  \bibinfo{author}{D.~Wackeroth}, \bibinfo{journal}{Phys. Rev.}
  \bibinfo{volume}{\textbf{D59}}, \bibinfo{pages}{013002}
  (\bibinfo{date}{1999}), \eprint{arXiv:hep-ph/9807417}.
\bibitem{Kripfganz:1988bd}
\bibinfo{author}{J.~Kripfganz} and \bibinfo{author}{H.~Perlt},
  \bibinfo{journal}{Z. Phys.} \bibinfo{volume}{\textbf{C41}},
  \bibinfo{pages}{319} (\bibinfo{date}{1988}).
\bibitem{Spiesberger:1995dm}
\bibinfo{author}{H.~Spiesberger}, \bibinfo{journal}{Phys. Rev.}
  \bibinfo{volume}{\textbf{D52}}, \bibinfo{pages}{4936} (\bibinfo{date}{1995}),
  \eprint{arXiv:hep-ph/9412286}.
\bibitem{dittm}
\bibinfo{author}{S.~Dittmaier} and \bibinfo{author}{M.~Kr{\"a}mer},
  \eprint{arXiv:hep-ph/0109062}.
\bibitem{Zykunov:2001mn}
\bibinfo{author}{V.~A. Zykunov}, \eprint{arXiv:hep-ph/0107059}.
\bibitem{BW}
\bibinfo{author}{U.~Baur} and \bibinfo{author}{D.~Wackeroth}, in preparation.
\bibitem{BKS}
\bibinfo{author}{U.~Baur}, \bibinfo{author}{S.~Keller}, and
  \bibinfo{author}{W.~K. Sakumoto}, \bibinfo{journal}{Phys. Rev.}
  \bibinfo{volume}{\textbf{D57}}, \bibinfo{pages}{199} (\bibinfo{date}{1998}),
  \eprint{arXiv:hep-ph/9707301}.
\bibitem{dg}
\bibinfo{author}{G.~Degrassi}, \bibinfo{author}{P.~Gambino}, and
  \bibinfo{author}{A.~Sirlin}, \bibinfo{journal}{Phys. Lett.}
  \bibinfo{volume}{\textbf{B394}}, \bibinfo{pages}{188} (\bibinfo{date}{1997}),
  \eprint{arXiv:hep-ph/9611363}.
\bibitem{BBHSW}
\bibinfo{author}{U.~Baur}, \bibinfo{author}{O.~Brein},
  \bibinfo{author}{W.~Hollik}, \bibinfo{author}{C.~Schappacher}, and
  \bibinfo{author}{D.~Wackeroth}, \eprint{arXiv:hep-ph/0108274}.
\bibitem{BS}
\bibinfo{author}{U.~Baur} and \bibinfo{author}{T.~Stelzer},
  \bibinfo{journal}{Phys. Rev.} \bibinfo{volume}{\textbf{D61}},
  \bibinfo{pages}{073007} (\bibinfo{date}{2000}),
  \eprint{arXiv:hep-ph/9910206}.
\bibitem{Gambino:1999ai}
\bibinfo{author}{P.~Gambino} and \bibinfo{author}{P.~A. Grassi},
  \bibinfo{journal}{Phys. Rev.} \bibinfo{volume}{\textbf{D62}},
  \bibinfo{pages}{076002} (\bibinfo{date}{2000}),
  \eprint{arXiv:hep-ph/9907254}.
\bibitem{Grassi:2001bz}
\bibinfo{author}{P.~A. Grassi}, \bibinfo{author}{B.~A. Kniehl}, and
  \bibinfo{author}{A.~Sirlin} (\bibinfo{date}{2001}),
  \eprint{arXiv:hep-ph/0109228}.
\bibitem{bardin}
\bibinfo{author}{D.~Y. Bardin}, \bibinfo{author}{A.~Leike},
  \bibinfo{author}{T.~Riemann}, and \bibinfo{author}{M.~Sachwitz},
  \bibinfo{journal}{Phys. Lett.} \bibinfo{volume}{\textbf{B206}},
  \bibinfo{pages}{539} (\bibinfo{date}{1988}).
\bibitem{Beenakker:1997kn}
\bibinfo{author}{W.~Beenakker} \emph{et~al.}, \bibinfo{journal}{Nucl. Phys.}
  \bibinfo{volume}{\textbf{B500}}, \bibinfo{pages}{255} (\bibinfo{date}{1997}),
  \eprint{arXiv:hep-ph/9612260}.
\bibitem{Argyres:1995ym}
\bibinfo{author}{E.~N. Argyres} \emph{et~al.}, \bibinfo{journal}{Phys. Lett.}
  \bibinfo{volume}{\textbf{B358}}, \bibinfo{pages}{339} (\bibinfo{date}{1995}),
  \eprint{arXiv:hep-ph/9507216}.
\bibitem{Baur:1995aa}
\bibinfo{author}{U.~Baur} and \bibinfo{author}{D.~Zeppenfeld},
  \bibinfo{journal}{Phys. Rev. Lett.} \bibinfo{volume}{\textbf{75}},
  \bibinfo{pages}{1002} (\bibinfo{date}{1995}), \eprint{arXiv:hep-ph/9503344}.
\bibitem{Passarino:1999zh}
\bibinfo{author}{G.~Passarino}, \bibinfo{journal}{Nucl. Phys.}
  \bibinfo{volume}{\textbf{B574}}, \bibinfo{pages}{451} (\bibinfo{date}{2000}),
  \eprint{arXiv:hep-ph/9911482}.
\bibitem{Ballestrero:2000kk}
\bibinfo{author}{A.~Ballestrero}, \bibinfo{journal}{Nucl. Phys. Proc. Suppl.}
  \bibinfo{volume}{\textbf{89}}, \bibinfo{pages}{25} (\bibinfo{date}{2000}),
  \eprint{arXiv:hep-ph/0005325}.
\bibitem{Beenakker:1999hi}
\bibinfo{author}{W.~Beenakker}, \bibinfo{author}{F.~A. Berends}, and
  \bibinfo{author}{A.~P. Chapovsky}, \bibinfo{journal}{Nucl. Phys.}
  \bibinfo{volume}{\textbf{B573}}, \bibinfo{pages}{503} (\bibinfo{date}{2000}),
  \eprint{arXiv:hep-ph/9909472}.
\bibitem{Denner:1996gb}
\bibinfo{author}{A.~Denner} and \bibinfo{author}{S.~Dittmaier},
  \bibinfo{journal}{Phys. Rev.} \bibinfo{volume}{\textbf{D54}},
  \bibinfo{pages}{4499} (\bibinfo{date}{1996}), \eprint{arXiv:hep-ph/9603341}.

\end{thebibliography}

\end{document}